\documentclass[%
 aip,
 amsmath,amssymb,
preprint,%
]{revtex4-1}

\usepackage{graphicx}
\usepackage{dcolumn}
\usepackage{bm}
\usepackage[mathlines]{lineno}

\usepackage[utf8]{inputenc}
\usepackage[T1]{fontenc}
\usepackage{mathptmx}
\usepackage{etoolbox}
\usepackage{textgreek}

\makeatletter
\def\@email#1#2{%
 \endgroup
 \patchcmd{\titleblock@produce}
  {\frontmatter@RRAPformat}
  {\frontmatter@RRAPformat{\produce@RRAP{*#1\href{mailto:#2}{#2}}}\frontmatter@RRAPformat}
  {}{}
}%
\makeatother
\begin{document}


\title[Quantum Cascade Surface Emitting Lasers]{Quantum Cascade Surface Emitting Lasers}

\author{David Stark}
\email[Author to whom correspondence should be addressed:]{ starkd@phys.ethz.ch}
\affiliation{Institute for Quantum Electronics, Department of Physics, ETH Z\"urich, 8093 Z\"urich, Switzerland}

\author{Filippos Kapsalidis}
\affiliation{Institute for Quantum Electronics, Department of Physics, ETH Z\"urich, 8093 Z\"urich, Switzerland}

\author{Sergej Markmann}
\affiliation{Institute for Quantum Electronics, Department of Physics, ETH Z\"urich, 8093 Z\"urich, Switzerland}

\author{Mathieu Bertrand}
\affiliation{Institute for Quantum Electronics, Department of Physics, ETH Z\"urich, 8093 Z\"urich, Switzerland}

\author{Bahareh Marzban}
\affiliation{Institute for Quantum Electronics, Department of Physics, ETH Z\"urich, 8093 Z\"urich, Switzerland}

\author{Emilio Gini}
\affiliation{FIRST Center for Micro- and Nanoscience, ETH Z\"urich, 8093 Z\"urich, Switzerland}

\author{Mattias Beck}
\affiliation{Institute for Quantum Electronics, Department of Physics, ETH Z\"urich, 8093 Z\"urich, Switzerland}

\author{Jérôme Faist}
\affiliation{Institute for Quantum Electronics, Department of Physics, ETH Z\"urich, 8093 Z\"urich, Switzerland}
\email{jfaist@ethz.ch}

\date{\today}

\begin{abstract}
    A low-cost single frequency laser emitting in the mid-infrared spectral region and dissipating minimal electrical power is a key ingredient for the next generation of portable gas sensors for high-volume applications involving chemical sensing of important greenhouse and pollutant gases.
    We propose here a Quantum Cascade Surface Emitting Laser (QCSEL), which we implement as a short linear cavity with high reflectivity coated end-mirrors to suppress any edge emission and use a buried semiconductor diffraction grating to extract the light from the surface. 
    By wafer-level testing we investigate the cavity length scaling, extract mirror reflectivities larger than 0.9, and achieve a pulsed threshold power dissipation of 237 mW for an emission wavelength near 7.5 \textmu m.
    Finally, we demonstrate single mode emission with a side-mode suppression ratio larger than 33 dB of a 248 \textmu m short cavity mounted with the epitaxial layer up and operated in continuous wave at 20 $^\circ$C.
    
\end{abstract}

\maketitle

\section{Introduction}
    The mid-infrared (MIR) spectral region spanning from 2 \textmu m to 20 \textmu m is the molecular “fingerprint” region for many important organic and inorganic molecules, as they exhibit strong and narrow absorption lines within this region \cite{Bernath2020SpectraMolecules,Gordon2017TheDatabase}.
    Miniaturized optical gas sensors based on MIR absorption spectroscopy are highly attractive \cite{Hodgkinson2012OpticalReview} for many applications such as industrial process control, environmental monitoring and medical diagnosis. \cite{Popa2019TowardsSensors}
    To enable low-cost and portable MIR gas sensors, compact and low power dissipation single mode light sources operating in the range of interest are of uttermost importance.

    The Quantum Cascade Laser (QCL) relying on intersubband transitions is an excellent candidate for a compact and coherent MIR light source because the emission wavelength can be tailored in wide ranges between 3 \textmu m and 24 \textmu m as well as between 60 \textmu m and 300 \textmu m. \cite{SerenaVitiello2015QuantumChallenges}
    The QCL can be modulated up to tens of GHz \cite{Hinkov2016Rf-modulationLasers} and exhibits narrow linewidths \cite{Argence2015QuantumLevel} making it the source of choice for fast high-resolution gas spectroscopy in the MIR. 
    
    However, harnessing intersubband transitions comes at the price of large threshold current densities ($\geq$ 0.47 kA/cm$^2$)\cite{Yan2015VeryLasers} which leads to an electrical power dissipation of several watts if a large device area is employed, hindering the integration into portable applications. 
    By scaling the cavity length the electrical power dissipation can be reduced if low optical losses can be maintained to achieve laser action in continuous wave (CW) above room-temperature.
    Particularly, high mirror reflectivities with values close to unity are essential to mitigate the increase of the mirror losses $\alpha_\text{m}$ with the reciprocal cavitity length $1/L$. This can be seen by considering the threshold current density $J_\text{th}$ given by
    \begin{equation} \label{eq:jth}
        J_\text{th} 
        = \frac{\alpha_\text{i} + \alpha_\text{m}}{g\Gamma}
        =
        \frac{1}{g\Gamma} \left(\alpha_\text{i} + \frac{\ln(1/R)}{L}\right),
    \end{equation}
    where $g$ is the material gain, $\Gamma$ the optical confinement factor, $\alpha_\text{i}$ the internal losses, and $R$ the mirror reflectivity. 
    Naturally, short cavity lasers not only enable smaller device footprints and thus a higher integration density, but they also offer a reduced number of axial lasing modes within the finite gain bandwidth and hence a route towards single axial mode selection. 
    While the Vertical Cavity Surface Emitting Laser (VCSEL) provides an excellent approach for interband lasers emitting in near-infrared spectral region \cite{2013VCSELs}, this approach cannot be used for QCLs because of the strict intersubband selection rule which requires the electric field of the optical mode to be perpendicular to the plane of the quantum wells.\cite{Faist2013QuantumLasers}

    Ridge QCLs with cavity lengths on the order of 100 \textmu m have been reported using deeply etched Bragg mirrors with estimated reflectivies larger than 0.8 \cite{Hofling2005DeviceBragg-mirrors}. An alternative approach relied on cleaving and depositing highly reflective (HR) metallic coatings with reflectivities of 0.95 and 0.75 on the front and back facet, respectively.\cite{Cendejas2011CavityOperation}
    For both approaches single mode operation could be observed, although the shortest devices were not operational at room-temperature.

    Superior CW temperature performance can be achieved with buried heterostructure QCLs\cite{Wittmann2009Distributed-feedbackK,Bismuto2015HighRange} featuring efficient thermal extraction \cite{Wang2020RoomOperation} and low waveguide losses $(0.5\;\text{cm}^{-1})$ \cite{Bai2011RoomEfficiency}.
    With a cavity length of 500 \textmu m and metallic HR (Al\textsubscript{2}O\textsubscript{3}/Ti/Au/Al\textsubscript{2}O\textsubscript{3}) coated back and partial HR (Al\textsubscript{2}O\textsubscript{3}/Ge) coated front facet, a threshold dissipation power of 260 mW at 10 $^\circ$C and 330 mW at 40 $^\circ$C has been reported. \cite{Cheng2020UltralowTemperature}
    Recently, a CW threshold dissipation power of only 143 mW at 20 $^\circ$C has been demonstrated with a cleaved 250 \textmu m short cavity. \cite{Wang2022Ultra-lowAperture} 
    This was achieved by fabricating a subwavelength aperture in the metallic HR (Al\textsubscript{2}O\textsubscript{3}/Au) coating on both facets to suppress diffraction losses.

    Those approaches, show that metallic HR coatings are convenient to shorten the cavity length, although both facets need to be coated and due to the low transmissivity of the metallic layer an aperture is required to still extract light. 
    To this end, buried second order diffraction gratings \cite{Yao2013Surface4.6m,Jouy2015SurfaceLasers} open up a way to couple out the light vertically. Furthermore, to scale the cavity length below 200 \textmu m \cite{Cendejas2011CavityOperation}, cleaving is not suitable, because the mechanical handling is challenging and the cavity length cannot precisely be controlled. Dry-etching controlled by lithography instead offers a promising alternative.
    
    Therefore, we introduce the Quantum Cascade Surface Emitting Laser (QCSEL), which is essentially an in-plane semiconductor laser implemented in a microcavity, combining an outcoupling element with HR mirrors to suppress any edge emission. 
    Here, we realize the QCSEL as a linear buried heterostructure laser with a buried second order grating in the vicinity of the waveguide, dry-etched facets and HR coatings deposited on wafer level. We demonstrate the operation of QCSELs with two different active regions designed for 4.5 \textmu m and 8 \textmu m emission wavelength. We utilize wafer-level testing to investigate the cavity length scaling to assess the mirror reflectivty and to ultimately reduce the power dissipation and the number of lasing modes.

\newpage

\section{Experiment}
    \begin{table}[b]
        \caption{\label{tab:sample_overview} 
        Overview of the fabricated quantum cascade active regions: Peak wavelength $\lambda_0$ and full width half maximum $2\gamma$ extracted from electroluminescence spectra, nominal sheet doping density $n_\text{2D}$, thickness of the top n-InGaAs cladding layer $t_\text{top}$, etch depth for the grating $d_\text{etch}$, and the thickness of the dielectric coating $t_\text{coat}$ extracted from scanning electron microscopy (SEM) cross-section images.}
        \begin{ruledtabular}
        \begin{tabular}{l |cc}
        Active region & EV1464 & EV2616 \\
        \hline
        $\lambda_0 \; (\mu\text{m})$ & 4.5 & 8 \\
        $2\gamma \; \big(\text{cm}^{-1}\big)$ & 238 & 198 \\
        $n_\text{2D} \big(10^{11}\text{cm}^{-2}\big)$  & 1.49 & 1.04 \\
        $t_\text{top} (\text{nm})$ & 200 & 300 \\
        $d_\text{etch} (\text{nm}) $ & 95 & 236 \\
        $t_\text{coat} (\text{nm}) $ & 435 & 300
        \end{tabular}
        \end{ruledtabular}
    \end{table}

    For this work we use two strain-balanced InGaAs/AlInAs active regions grown on InP substrates by molecular beam epitaxy (MBE).
    From electroluminescence measurements, the peak wavelength and the full width half maxima estimating the gain bandwidth are inferred for both active regions. These values together with the sheet doping densities and the relevant fabrication parameters discussed below are summarized in Table~\ref{tab:sample_overview}.
    
\begin{figure*}[t]
        \includegraphics[width=1\linewidth]{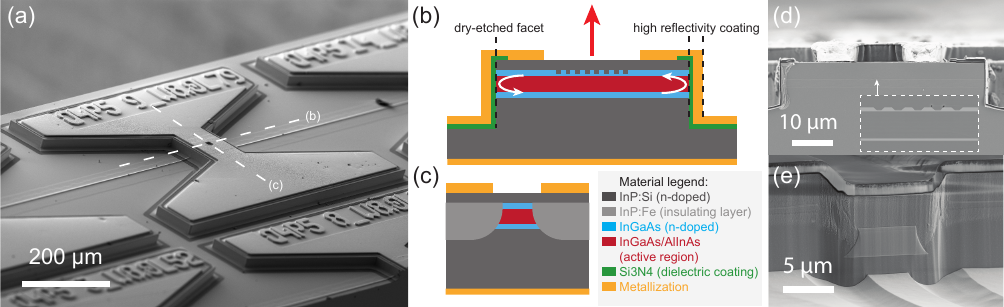}
        \caption{\label{fig:concept} The QCSEL device architecture: 
        (a) Scanning electron microscopy (SEM) image of the device after fabrication. 
        The device footprint is below 500 $\times$ 400 \textmu m$^2$ and the cavity length is 79 \textmu m.
        The light is extracted through the metallic aperture in the center of the device indicated by the crossing of the two dashed lines.
        The dashed lines display the axial and lateral direction. The schematic diagrams in (b) and (c) show the axial and the lateral cross-section of the linear cavity, respectively. 
        The red arrow indicates the vertical light extraction. 
        (d) SEM of a cross-section of a 50 \textmu m short linear cavity corresponding to schematic diagram in (b). The inset displays a magnified view of the grating section where the InGaAs layers are highlighted for clarity.
        (e) SEM image of a dry-etched facet of a test structure showing the etch quality.
        }
    \end{figure*}

    The buried heterostructure fabrication process \cite{Beck2002ContinuousTemperature,Suess2016AdvancedMIR-QCLs} starts with the definition and wet-etching of the second order grating on the top n-InGaAs cladding layer. The waveguide core is also formed by wet-etching. The lateral semi-insulating InP:Fe blocking layer and the top InP:Si contact layer are regrown by means of metal-organic vapor phase epitaxy (MOVPE).  
    After depositing the metallic top-contact (Ti/Pt/Au), the end-mirror facets are dry-etched with inductively coupled plasma (ICP).
    Subsequently, both facets are coated with Si\textsubscript{3}N\textsubscript{4}/Ti/Au to build HR end-mirrors. 
    The former dielectric layer is deposited by plasma-enhanced chemical vapor deposition (PECVD) and the latter metallic layers by electron beam evaporation. 
    Note that for the metallic coating a Ti adhesion layer before the Au deposition is necessary for further processing and we estimate its thickness to be thinner than 20 nm.
    After revealing the top-contact by wet- and dry-etching, an additional Au layer with a thickness between 3 \textmu m to 5 \textmu m is electroplated. 
    Lastly, the substrate is thinned down to about 200 \textmu m and a back contact (Ge/Au/Ni/Au) is deposited. 
    The QCSEL device architecture is illustrated in Fig.~\ref{fig:concept} with fabrication images and cross-section schematics. 

\begin{figure}[h]
        \includegraphics{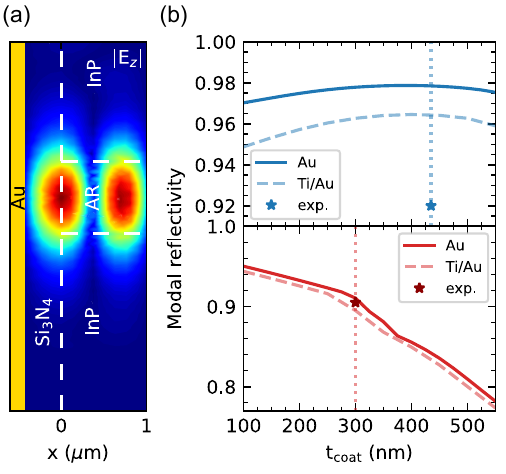}
        \caption{\label{fig:reflectivity} 
        3D Simulation of the facet reflectivity:
        (a) Electric field pattern along the growth direction $|\mathbf{E}_z(x,z)|$ for 1 \textmu m waveguide section terminated with a Si\textsubscript{3}N\textsubscript{4}/Au coating. 
        Note that a rectangular waveguide is assumed with constant width. Here, the wavelength is 4.5 \textmu m, the active region thickness is 2.03 \textmu m, and $t_\text{coat} = 435\;\text{nm}$.
        (b) Modal reflectivity for varying $t_\text{coat}$ and wavelengths of 4.5 \textmu m (top panel) and 8 \textmu m (bottom panel). The dashed lines include a 20 nm thick Ti adhesion layer.
        The vertical dotted lines correspond to $t_\text{coat}$ of the fabrication (see Table~\ref{tab:sample_overview}) and the stars indicate the experimentally deduced reflectivies discussed below.
        }
    \end{figure}

    To estimate the refelectivity of the HR end-mirrors, we performed 3D simulations (COMSOL) of a 1 \textmu m waveguide section terminated with a Si\textsubscript{3}N\textsubscript{4}/Au coating, see Fig.~\ref{fig:reflectivity}(a). 
    The modal reflectivities for varying dielectric coating thickness $t_\text{coat}$ with and without Ti adhesion layer are shown in Fig.~\ref{fig:reflectivity}(b).
    For a wavelength of 4.5 \textmu m and 8 \textmu m, we expect a modal reflectivity larger than 0.96 and 0.89, respectively.
    At a wavelength of 8 \textmu m the absorption losses of $\text{Si}_3\text{N}_4$ are dominating \cite{Kischkat2012Mid-infraredNitride} and although thinner coatings are more beneficial, we chose $t_\text{coat}=300\;\text{nm}$ to ensure the electrical insulation of the coating.
    
    In this work, the active device length is scaled from 504 \textmu m down to 46 \textmu m. Approximating the cavity length $L$ by the active device length and neglecting the optical path length in the HR coatings ($L\gg t_\text{coat}$), the free spectral range $\Delta\nu$ can be written as
    \begin{equation} \label{eq:fsr}
        \Delta \nu = \frac{1}{2n_\text{g} L},
    \end{equation}
    where $n_\text{g}$ is the group index of the guided mode in the active region.
    By further assuming $n_\text{g}=3.4$, $\Delta\nu$ ranging from 3 cm$^{-1}$ to 32 cm$^{-1}$ are accessible in our experiments.

\begin{figure}[t]
        \includegraphics{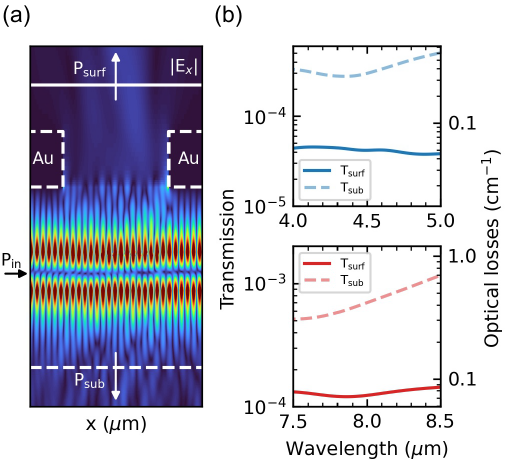}
        \caption{\label{fig:extraction} 
        2D Simulation of the outcoupling element using 7 grating periods:
        (a) Electric field pattern along the waveguide $|\mathbf{E}_x(x,z)|$ for $\lambda=8\;\mu\text{m}$.
        The input power $P_\text{in}$ exciting the axial waveguide mode and the monitors detecting the power directed towards the surface $P_\text{surf}$ and the substratte $P_\text{sub}$ are illustrated.
        (b) The transmission $T$ towards the surface $T_\text{surf}=P_\text{surf}/P_\text{in}$ and the substrate $T_\text{sub}=P_\text{sub}/P_\text{in}$ versus the wavelength. The optical losses of a single pass are computed using Eq.~(\ref{eq:grating_losses}).
        The grating etch depths and the thicknesses of the top n-InGaAs cladding layers of Table~\ref{tab:sample_overview} are used.
        }
    \end{figure}

    The design of the outcoupling element, a non-resonant diffraction grating, follows the approach by Jouy and co-workers. \cite{Jouy2015SurfaceLasers} 
    The near-second order grating period is determined by
    \begin{equation}
        \Lambda = \frac{N-0.5}{N}\cdot \frac{\lambda}{n_\text{eff}}.
    \end{equation}
    where $N$ is the number of grating periods, $n_\text{eff}$ the effective guided mode index, and $\lambda$ the free space wavelength.
    The outcoupling is investigated with 2D simulations (Lumerical) where the length of the grating section $L_\text{gs}$ is fixed by $N=7$ and the fabrication parameters of the respective active region are used (see Table~\ref{tab:sample_overview}).
    The simulation and the electric field component responsible for the vertical outcoupling for a wavelength of $8\;\mu\text{m}$ are illustrated in Fig.~\ref{fig:extraction}(a).
    The transmission towards the surface and the substrate for varying wavelengths and the corresponding induced optical losses of a single pass of the grating section are considered in Fig.~\ref{fig:extraction}(b).
    The optical losses are expressed in terms of the length of the grating section $L_\text{gs}$ using
    \begin{equation} \label{eq:grating_losses}
        \alpha = -\frac{\ln(1-T)}{L_\text{gs}}.
    \end{equation}
    Here, $T$ is the transmission, which corresponds to either the transmission to the surface, the transmission to the substrate, or the transmission back to the input (reflection, not shown in Fig.~\ref{fig:extraction}(b)). 
    For both wavelength ranges, the surface extraction losses are $\alpha_\text{surf} < 0.1\;\text{cm}^{-1}$, the substrate losses $\alpha_\text{sub} > 0.3\;\text{cm}^{-1}$, and the reflection losses $\alpha_\text{refl} < 0.04\;\text{cm}^{-1}$.
    While for long cavities $(L\gg L_\text{gs})$, such an outcoupling element results in small slope efficiencies ($\eta\propto \alpha_\text{surf}/\alpha_\text{tot}$, where $\alpha_\text{tot}$ are the total optical losses), an enhancement of the slope efficiency is expected as the cavity length is scaled $(L\sim L_\text{gs})$. 
    This is because the induced surface extraction losses can be seen as localized mirror losses which are proportional to the reciprocal length. Besides the short grating section, we also use small metallic apertures $(\leq 21.2 \times 10.5 \;\mu\text{m}^2)$ to ensure uniform electrical pumping and sufficient thermal extraction.
    
    To characterize the QCSELs we developed an automatized probe-station to perform light-current-voltage (LIV) characterization on wafer-level, where the individual QCSELs can be adressed with a motorized stage.
    Light is collected and refocused on the detector with two 3 in off-axis parabolic mirrors, where the first mirror has a numerical aperture of about 0.7. 
    Either a HgCdTe detector (Vigo systems, model: PVM-2TE-10.6-1x1-TO8-wZnSeAR-70+MIP-DC-250M) or a power meter are used as a detector.
    For spectral and CW characterization some devices were mounted on submounts.
    Initially the QCSELs are characterized in pulsed operation with a repetition rate of $96.15\;\text{kHz}$ and a pulse width of $312\;\text{ns}$.
    Also cleaved reference lasers without any coating and diffraction grating are characterized to estimate the material gain (discussed below) and the average widths of the QCSELs which are obscured by the HR coatings (see Fig.~\ref{fig:concept}).
    The average widths are computed by fitting the sub-threshold IV-curves of the QCSELs to a sub-threshold IV-curve of a reference laser.
    Using these widths, instead of individual widths inferred during the fabrication, more accurate current density values for QCSELs across the whole sample can be obtained.
    All the measurements presented in the following section are performed at 20$^\circ$C.

\clearpage   
\section{\label{sec:level1} Results}
    \begin{figure*}[b]
        \includegraphics[width=0.9\linewidth]{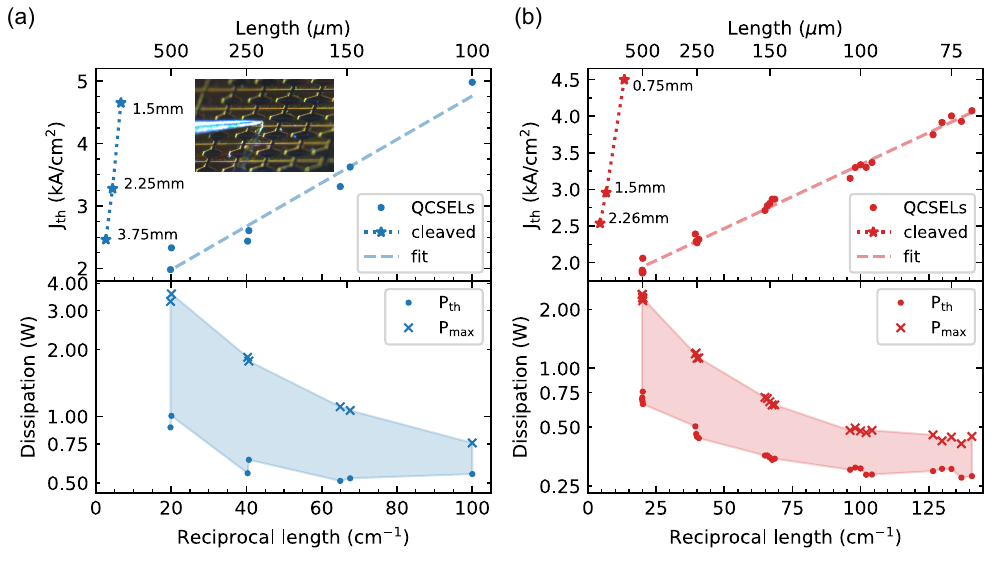}
        \caption{\label{fig:dissipation_jth} 
        Cavity length scaling based on the active region EV1464 (a) and EV2616 (b). 
        The top panels show the threshold current densities over the reciprocal length for cleaved reference lasers without extractor and the QCSELs. The lengths of the reference lasers are annotated. The bottom panels show the corresponding electrical power dissipation of the QCSELs in a semilogarithmic plot.
        Note that the device lengths of the QCSELs are indicated with the top axes in each panel.
        The inset of the top panel (a) shows the probing of the QCSELs on wafer-level.
        }
    \end{figure*}
    As mentioned above, to exploit the advantages of short cavity lasers, i.e. single frequency operation at minimal electrial power dissipation on a compact device footprint, high facet reflectivities with values close to unity are required.
    We estimate these facet reflectivities by measuring the threshold of lasers with varying lengths.
    The threshold current densities $J_\text{th}$ of QCSELs and reference lasers are extracted and shown in the top panels of Fig.~\ref{fig:dissipation_jth}. 
    As expected from Eq.~(\ref{eq:jth}), $J_\text{th}$ increases linearly with the reciprocal length.
    The material gain $g\Gamma$ is estimated using Eq.~(\ref{eq:jth}) and fitting $J_\text{th}$ of the reference lasers with cleaved facets for which we assume equal reflectivities of 0.28. 
    Finally, with $g\Gamma$, $J_\text{th}$ of the QCSELs and Eq.~(\ref{eq:jth}), the reflectivity of the QCSEL facets is deduced.
    These results are summarized in Table~\ref{tab:results_reciprocal_length} and it can be seen that the experimental reflectivity of 0.905 agrees well with the simulations for EV2616.
    Albeit a higher experimental reflectivity of 0.92 is obtained for EV1464, the value deviates from the simulated values by more than $0.03$.
    This discrepancy could originate from the roughness induced by dry-etching, which is neglected in our simulations. 
    Due to the shorter emission wavelength of EV1464 compared to EV2616, we expect the reflectivity of the end-mirrors to be more prone to roughness.
    Considering the simulations shown in Fig.~\ref{fig:reflectivity}(b), at a wavelength of 4.5 \textmu m the reflectivity can be improved potentially by more than 0.01. This improvement can be achieved by minimizing the thickness of the Ti adhesion layer which ultimately reduces the mirror losses by more than 28 \%.
    At a wavelength of 8 \textmu m the refelectivity can be additionally increased to about 0.95 by reducing $t_\text{coat}$ and another 0.01 improvement is predicted by employing Al\textsubscript{2}O\textsubscript{3} as dielectric coating. 
    Lastly, for both wavelengths the diffraction losses at the facet can still be reduced by adjusting the refractive index profile of the waveguide core e.g. by changing the thickness of the active region or the cladding layers.

    \begin{table}[t]
        \caption{\label{tab:results_reciprocal_length} 
        Reciprocal length study for both active regions: 
        Experimental material gain $g\Gamma$, experimental reflectivity $R$, and the simulated reflectivity with and without the Ti adhesion layer $\widetilde{R}_\text{Ti/Au}$ and $\widetilde{R}_\text{Au}$, respectively. Note that for the cleaved facets a reflectivity of $0.28\pm 0.01$ is assumed.
        }
        \begin{ruledtabular}
        \begin{tabular}{ccccc}
        Active region & $g\Gamma$ (cm/kA) & $R$ & $\widetilde{R}_\text{Ti/Au}$ & $\widetilde{R}_\text{Au}$ \\
        \hline
        EV1464 & $2.3\pm 0.2$ & $0.92\pm 0.01$ & 0.964 & 0.978\\
        EV2616 & $5.7\pm 0.3$ & $0.905 \pm 0.005 $& 0.895 & 0.910
        \end{tabular}
     \end{ruledtabular}
    \end{table}
    
\begin{figure*}[h]
        \includegraphics[width=0.5\linewidth]{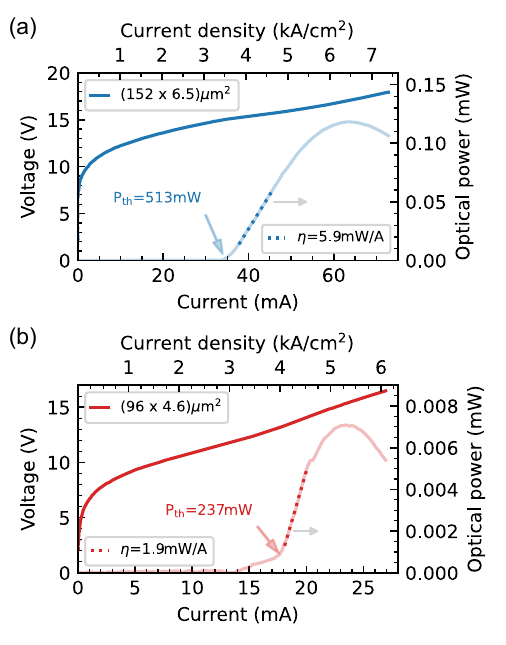}
        \caption{\label{fig:LIV_low_Pth} 
        Pulsed light-current-voltage (LIV) characteristics with the lowest power dissipation at threshold $P_\text{th}$ for QCSELs based on EV1464 (a) and EV2616 (b).
        The device area and the slope efficiencies $\eta$ are annotated.
        }
    \end{figure*}

    The scaling of the electrical power dissipation versus the reciprocal length is shown in the bottom panels of Fig~\ref{fig:dissipation_jth}. 
    The filled area illustrates the overall power dissipation, the limiting lower and upper curves indicate the dissipation at threshold $P_\text{th}$ and at maximum optical power $P_\text{max}$, respectively. 
    For a QCSEL based on EV1464 and an active area of $(100\times 6.8)$ \textmu m$^2$, $P_\text{max}$ and $P_\text{th}$ are reduced to 758 mW and 548 mW, respectively.
    For a QCSEL based on EV2616 and an active area of $(73\times 7.1)$ \textmu m$^2$, $P_\text{max}$ and $P_\text{th}$ are reduced to 411 mW and 276 mW, respectively.
    Among all the characterized QCSELs ($>240$ lasers), the pulsed LIV-characteristics of QCSELs with lowest $P_\text{th}$ for both active regions are reported in Fig.~\ref{fig:LIV_low_Pth}. 
    Compared to EV1464 where a $P_\text{th}$ of 513 mW is observed, a much lower $P_\text{th}$ of 237 mW could be achieved for EV2616 with the lower sheet doping density.
    The overall power dissipation can easily be improved by decreasing the doping levels in the claddings and the active region and by further employing a narrow gain active region. 
    \newpage
    
\begin{figure*}
        \includegraphics[width=0.8\linewidth]{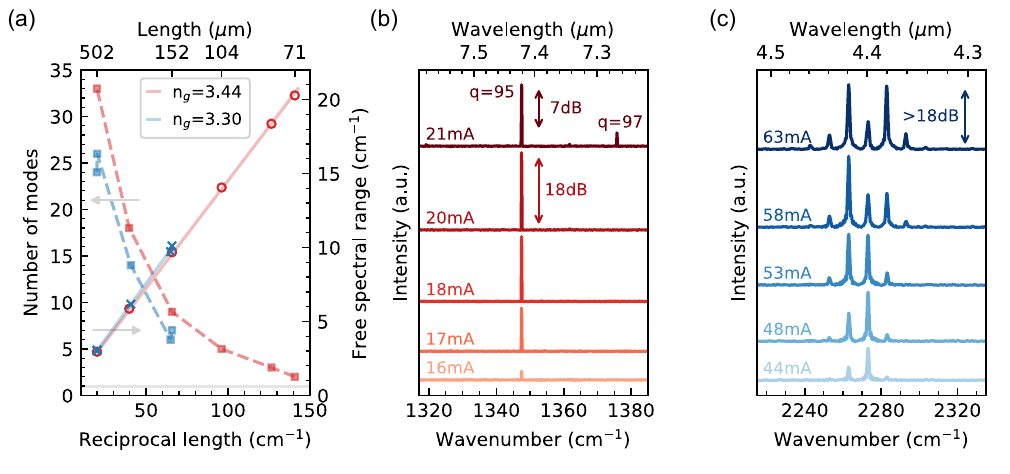}
        \caption{\label{fig:spectra} 
        Pulsed spectral characterization of the QCSELs:
        (a) Number of modes (square markers) and free spectral ranges (circle and cross markers) extracted from spectra acquired for currents close to rollover. 
        The solid lines correspond to the fit of the free spectral range (see Eq.(\ref{eq:fsr})).
        The horizontal gray line indicates the number of modes equals one. 
        (b) Spectra acquired from a QCSEL with $L=104$ \textmu m.
        The axial mode index $q$ and the side mode suppression ratio are illustrated.
        (c) Spectra acquired from a QCSEL with $L=152$ \textmu m (corresponding LIV shown in Fig.~\ref{fig:LIV_low_Pth}(a)). The arrow indicates the signal-to-noise ratio.
        Note that all curves appearing blueish are acquired from QCSELs based on EV1464 (wafer-level) with $0.5 \;\text{cm}^{-1}$ resolution and a pulse width of $52\;\text{ns}$. 
        All curves appearing reddish are acquired from QCSELs based on EV2616 (chip mounted on submounts) with $0.075\;\text{cm}^{-1}$ and a pulse width of $52\;\text{ns}$. 
        }
    \end{figure*}

    In Fig.~\ref{fig:spectra}(a), we demonstrate the reduction of axial lasing modes towards single mode operation through cavity length scaling.
    For QCSELs with varying lengths, the number of modes and the corresponding free spectral range were assessed from spectra acquired for currents close to rollover.
    The free spectral range is then fitted to Eq.~(\ref{eq:fsr}) resulting in $n_\text{g}=3.30$ and $n_\text{g}=3.44$ for EV1464 and EV2616, respectively.
    Although for the shortest device ($L=71$ \textmu m) still two modes are observed, this is an encouraging result because the active region is originally designed for broad gain employing two active region stacks. Moreover, the device lengths are not designed such that peak of the gain curve is matching an axial cavity mode.
    Thus, we are convinced that reliable single axial mode selection can be achieved with a narrow gain and single stack active region featuring a vertical optical transition.
    Nevertheless, a dominant single mode could be observed with a side mode suppression ratio (SMSR) of 18 dB at 20 mA from a QCSEL based on EV2616 ($L=104$ \textmu m), see Fig.~\ref{fig:spectra}(b).
    This mode is centered at 1347.5 cm$^{-1}$ with the axial mode index $q=95$ and for larger currents the SMSR decreases as the intensity of the mode $q=97$ increases.
    For a QCSEL based on the shorter wavelength active region EV1464 ($L=152$ \textmu m), the spectra for varying currents are shown in Fig.~\ref{fig:spectra}(c) and
    multiple modes around 4.4 \textmu m are observed. Compared to the spectra of the QCSEL shown in Fig.~\ref{fig:spectra}(b), more modes show up due to the longer cavity length and therefore smaller free spectral range. Additionally, from the electroluminescence measurements (see Table~\ref{tab:sample_overview}) we estimate a larger gain bandwidth for the active region EV1464.
    
    \begin{figure}[b]
        \includegraphics{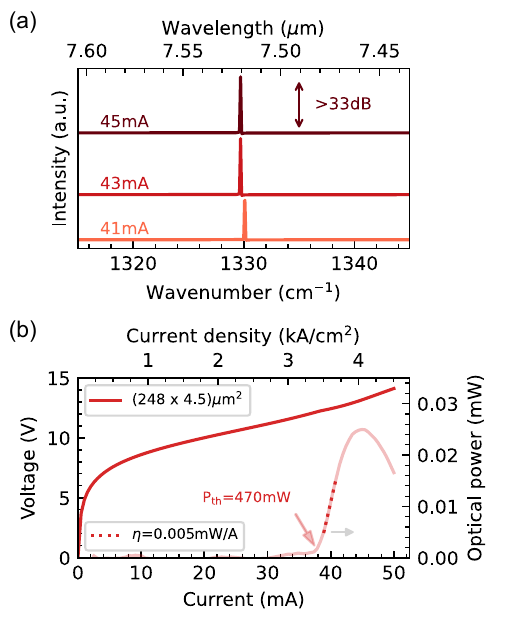}
        \caption{\label{fig:CW}
        Continuous wave characterization of a QCSEL based on EV2616 (chip mounted on submount): 
        (a) Spectra acquired with a resolution of $0.075 \;\text{cm}^{-1}$. The side mode suppression reatio is illustrated for a current of 45 mA.
        (b) LIV characteristics with annotated device area, slope efficiency $\eta$ and threshold dissipation power $P_\text{th}$.
        }
    \end{figure}
    Single mode emission can be observed when operating the QCSELs in CW.
    The spectra and the corresponding LIV for a QCSEL based on EV2616 ($L=248$ \textmu m) is shown in Fig.~\ref{fig:CW}.
    Although the mode selection is not guaranteed, only one mode appears and at $45\;\text{mA}$ a SMSR larger than $33\;\text{dB}$ is deduced.
    It needs to be emphasized, that the chip used for these CW measurements is mounted on a submount with epitaxial layer up and that improvements are expected from a thinner substrate or from a double-channel geometry \cite{Cheng2020UltralowTemperature}.
    
\begin{figure*}[h]
        \includegraphics[width=0.8\linewidth]{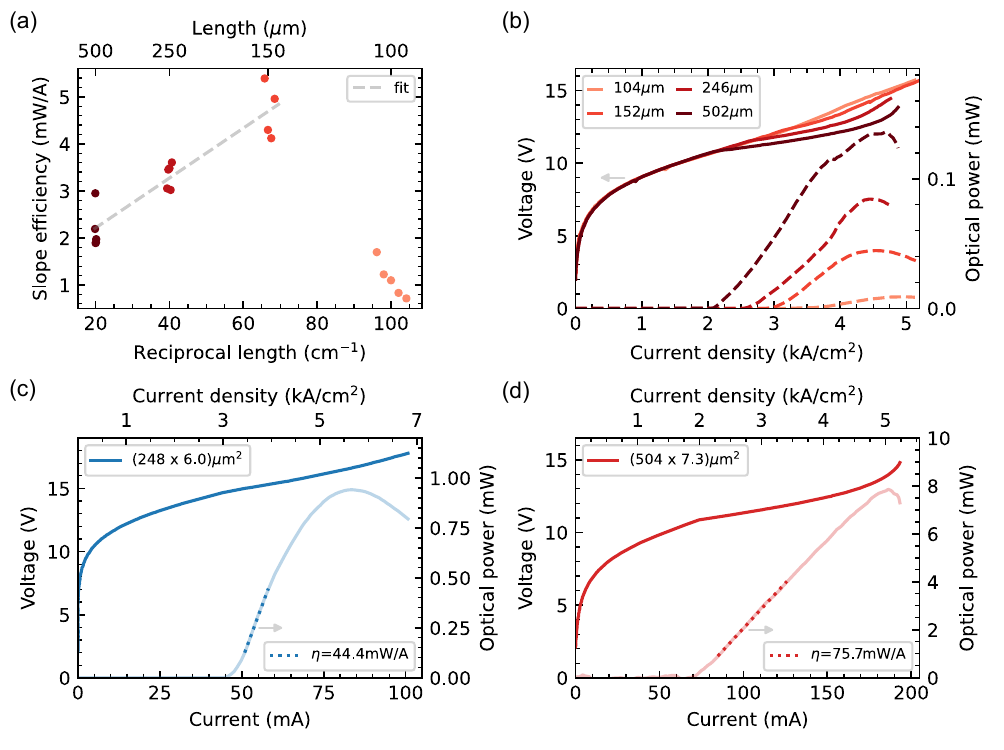}
        \caption{\label{fig:slope} 
        (a) Slope efficiency versus the reciprocal length extracted from LI curves of QCSELs based on EV2616. (b) Pulsed LIV characteristics for a selection of QCSELs used in (a). The device lengths are annotated.
        (c)-(d) Pulsed LIV characteristics exhibiting the largest slope efficiencies for QCSELs based on EV1464 (c) and EV2616 (d). The device areas and the slope efficiencies $\eta$ are annotated.
        }
    \end{figure*}

    An enhancement of the slope efficiency as discussed above is demonstrated in Fig.~\ref{fig:slope}(a), showing that for $L\geq 146$ \textmu m the slope efficiency increases roughly linearly with the reciprocal length as expected. 
    For smaller $L$ the slope efficiency then drops because $J_\text{th}$ increases close to the maximum current density $J_\text{max}$, which can be seen from the corresponding LIV-characteristics shown in Fig.~\ref{fig:slope}(b).
    To extend the enhancement of the slope efficiency for shorter cavities, higher reflectivities of the end-mirrors are required. 
    In Fig.~\ref{fig:slope}(c)-(d), the pulsed LIV-characteristics of QCSELs exhibiting the largest slope efficiencies are reported.
    The values for the slope efficiencies are $44.4\;\text{mW/A}$ and $75.7\;\text{mW/A}$ for EV1464 and EV2616, respectively. 
    These exceptionally higher values might be explained by a more favorable alignment between the dry-etched end-mirrors or by unintentional air voids during the regrowth step after wet-etching the grating. 
    Creating instead air voids on purpose by a two-step regrowth, employing a thicker etch step into the n-InGaAs cladding layer, or extending the area of the metallic aperture provide ways to increase the surface extraction losses and ultimately the slope efficiency.

\clearpage
\newpage

\section{Conclusions}
    We have proposed the QCSEL for the next generation of portable and large scale applications relying on gas phase absorption spectroscopy. 
    The QCSEL was implemented as a compact buried heterostructure laser with footprints well below $(500\times400)$ \textmu m$^2$ based on two different active regions designed for 4.5 \textmu m and 8 \textmu m.
    Lasing is demonstrated for devices as short as 71 \textmu m showing that higher integration densities are feasible.
    The key to further down scale the cavity length accompanied by the reduction of the electrical power dissipation and the number of axial lasing modes, is to increase the mirror reflectivites above 0.92 achieved in this work. 
    This can be done by leveraging on high-quality dry-etching, minimizing the Ti adhesion layer, and adjusting the refractive index profile of the waveguide core to reduce diffraction losses.
    To reduce the power dissipation beyond 237 mW and to guarantee single axial mode selection, low doped and narrow gain active regions will be combined with cavity lengths designed such that a single axial mode is matching the peak wavelength of the gain curve.

\section{Acknowledgements}
    The authors gratefully acknowledge the financial support from Innosuise - Swiss Innovation Agency (Innovation Projects: 52899.1 and 53098.1) and the ETH Zürich Foundation (Project: 2020-HS-348).
    \\
    The authors express their gratitude to Zhixin Wang and Ruijung Wang for useful discussion and conceiving ideas.
    Furthermore, the authors would like to thank Bo Meng for helpful hints for the device fabrication, Moritz Müller for developing further the automatized probe-station, and Philipp Täschler for valuable inputs on the manuscript text.

\section{Data availability}
    The data that support the findings of this study are available from the corresponding author upon reasonable request.

\section{Keywords}
    quantum cascade lasers, mid-infrared, single mode, low dissipation, surface emission, scaling, microcavity

\clearpage
\newpage
\nocite{*}
\bibliography{main}

\end{document}